\documentclass[aip,
 mph,%
 amsmath,
 amssymb,
 superscriptaddress,
 floatfix,
 preprint,
 eqsecnum,%
 numerical,%
]{revtex4-1}

\usepackage{graphicx}
\DeclareGraphicsExtensions{.eps,.png,.jpg}
\usepackage[caption=false
]{subfig}
\usepackage{epsfig}
\usepackage{epstopdf}
\usepackage{lipsum}
\usepackage{wrapfig}
\usepackage{wasysym}
\usepackage{threeparttable,booktabs,siunitx}
\usepackage{booktabs}
\usepackage{lmodern}
\usepackage{longtable}
\usepackage{multirow}
\usepackage{amsmath}
\usepackage[OT2,T1]{fontenc}
\usepackage[bookmarks=false]{hyperref}
\usepackage{color}
\hypersetup{pdfauthor={},pdfborder=0 0 0,colorlinks=true,citecolor=blue,linkcolor=blue,urlcolor=blue,}

\DeclareSymbolFont{cyrletters}{OT2}{wncyr}{m}{n}
\DeclareMathSymbol{\comb}{\mathalpha}{cyrletters}{"58}

\newcommand{\ben}{\begin{eqnarray}\displaystyle}
\newcommand{\een}{\end{eqnarray}}

\begin{document}

\title{DIRECT-Net: a unified mutual-domain material decomposition network for quantitative dual-energy CT imaging}

\author{Ting Su}
\affiliation{Research Center for Medical Artificial Intelligence, Shenzhen Institutes of Advanced Technology, Chinese Academy of Sciences, Shenzhen, Guangdong 518055, China.}
\author{Xindong Sun}
\affiliation{School of Information and Communication Engineering, University of Electronic Science and Technology of China, Chengdu 611731, China.}
\author{Yikun Zhang}
\affiliation{School of Computer Science and Engineering, Southeast University, Nanjing, Jiangsu 210096, China.}%
\author{Haodi Wu}
\affiliation{Wuhan National Laboratory for Optoelectronics (WNLO), Huazhong University of Science and Technology (HUST), 430074 Wuhan, China.}%
\author{Jianwei Chen}
\affiliation{Research Center for Medical Artificial Intelligence, Shenzhen Institutes of Advanced Technology, Chinese Academy of Sciences, Shenzhen, Guangdong 518055, China.}
\author{Jiecheng Yang}%
 \affiliation{Research Center for Medical Artificial Intelligence, Shenzhen Institutes of Advanced Technology, Chinese Academy of Sciences, Shenzhen, Guangdong 518055, China.}
\author{Yang Chen}
\affiliation{School of Computer Science and Engineering, Southeast University, Nanjing, Jiangsu 210096, China.}%
\author{Hairong Zheng}%
\affiliation{Paul C Lauterbur Research Center for Biomedical Imaging, Shenzhen Institutes of Advanced Technology, Chinese Academy of Sciences, Shenzhen, Guangdong 518055, China.}%
\author{Dong Liang}%
 \affiliation{Research Center for Medical Artificial Intelligence, Shenzhen Institutes of Advanced Technology, Chinese Academy of Sciences, Shenzhen, Guangdong 518055, China.}
\affiliation{Paul C Lauterbur Research Center for Biomedical Imaging, Shenzhen Institutes of Advanced Technology, Chinese Academy of Sciences, Shenzhen, Guangdong 518055, China.}%
\affiliation{University of Chinese Academy of Sciences, Beijing, 100049, China.}%
\author{Yongshuai Ge}%
 \email{ys.ge@siat.ac.cn.}
 \affiliation{Research Center for Medical Artificial Intelligence, Shenzhen Institutes of Advanced Technology, Chinese Academy of Sciences, Shenzhen, Guangdong 518055, China.}
\affiliation{Paul C Lauterbur Research Center for Biomedical Imaging, Shenzhen Institutes of Advanced Technology, Chinese Academy of Sciences, Shenzhen, Guangdong 518055, China.}%
\affiliation{University of Chinese Academy of Sciences, Beijing, 100049, China.}%

\date{\today}

\begin{abstract}
By acquiring two sets of tomographic measurements at distinct X-ray spectra, the dual-energy CT (DECT) enables quantitative material-specific imaging. However, the conventionally decomposed material basis images may encounter severe image noise amplification and artifacts, resulting in degraded image quality and decreased quantitative accuracy. Iterative DECT image reconstruction algorithms incorporating either the sinogram or the CT image prior information have shown potential advantages in noise and artifact suppression, but with the expense of large computational resource, prolonged reconstruction time, and tedious manual selections of algorithm parameters. To partially overcome these limitations, we develop a domain-transformation enabled end-to-end deep convolutional neural network (DIRECT-Net) to perform high quality DECT material decomposition. Specifically, the proposed DIRECT-Net has immediate accesses to mutual-domain data, and utilizes stacked convolution neural network (CNN) layers for noise reduction and material decomposition. The training data are numerically simulated based on the underlying physics of DECT imaging.The XCAT digital phantom, iodine solutions phantom, and biological specimen are used to validate the performance of DIRECT-Net. The qualitative and quantitative results demonstrate that this newly developed DIRECT-Net is promising in suppressing noise, improving image accuracy, and reducing computation time for future DECT imaging.
\end{abstract}

\maketitle

\section{Introduction}
\label{sec:introduction}
Spectral computed tomography (CT) has attracted many research interests for clinical applications. By adopting certain data acquisition hardware and material decomposition algorithms, quantitative material-specific imaging can be achieved. As one special case, the dual-energy CT (DECT) imaging technique takes the measurements at two different X-ray spectra. Due to its superior material discrimination ability, DECT has been widely used for a variety of medical applications, including kidney stone characterization, gout diagnosis, contrast agent enhanced lesion detection, and so on \cite{Patio2016MaterialSU}. In spite of the clinical advancement, unfortunately, there still remain some challenges in DECT imaging. For instance, most of the currently used material decomposition algorithms for DECT imaging usually lead to strong noise amplification \cite{Zhao2016UsingEA}, and thus degrade the signal-to-noise ratio (SNR).

\begin{figure}[htb]
	\centering
	\includegraphics[width=1\textwidth]{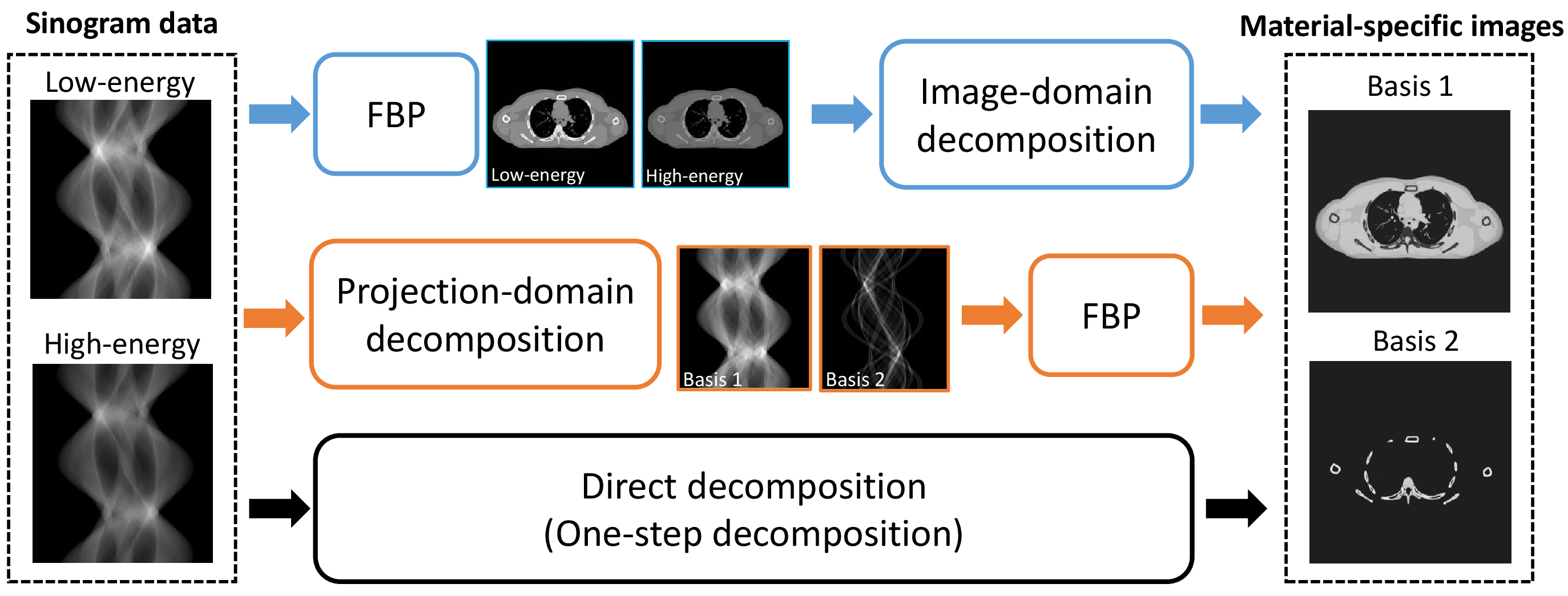}
	\caption{Three major strategies to perform DECT image decomposition. From top to bottom, they are: the image-domain method, the projection-domain method, and the direct method.}
	\label{fig_decomp}
	\end{figure}

In general, the material decomposition algorithms can be categorized into three groups: the image-domain decomposition method, the projection-domain decomposition method, and the so-called direct decomposition or one-step decomposition method using mutual-domain information, as schemed in Fig.~\ref{fig_decomp}. For each scheme, there are pros and cons. (1) The image-domain decomposition algorithm reconstructs the dual-energy CT images from the measured sinogram data at first, and then performs decomposition directly on the CT images using linear approximations. Various image domain prior information can be incorporated, such as the total variation (TV) \cite{doi:10.1118/1.4870375}, nonlocal TV \cite{7506078}, edge-preserving quadratic smoothness penalty \cite{Niu2014IterativeID}, sparsity and low-rank property\cite{Gao_2011}, non-negativity constraint \cite{6600785},  spatial spectral non-local means \cite{Li2018MultienergyCC}, prior image constrained compressed sensing (PICCS) \cite{Szczykutowicz_2010}, etc. These image-domain interative algorithms are fairly easy to be implemented. However, artifacts such as the beam hardening effect are hard to be mitigated. (2) The projection-domain algorithms firstly estimate the material-specific projections and then generate the basis CT images using conventional reconstruction algorithm \cite{Alvarez1976EnergyselectiveRI, Schlomka_2008, Su2018ASX}. The statistical noise model of the measured projection data can be exploited to improve the SNR, and the beam hardening effects can be eliminated as well \cite{6473891, PMID:29570809}. However, the decomposition performance depends heavily on the accuracy of estimated spectra, and the spacial prior information in image domain are also not fully utilized. (3) The direct decomposition algorithms (also known as the one-step algorithms) depict the DECT image reconstruction process with a complicated non-linear data model that takes the material-specific maps as variables, and the unknown basis images are solved iteratively by minimizing a certain objective function \cite{Mechlem2018JointSI, Foygel_Barber_2016, Schmidt2017ASC, 2013MedPh..40k1916C, Long2014MultiMaterialDU, Mory2018ComparisonOF}. The direct algorithms have some advantages over the two aforementioned algorithms, because both the projection-domain and CT image-domain prior information are used simultaneously to yield improved image quality and accuracy. Nevertheless, besides the heavy spectra dependence problem, forward and back projections have to be performed repeatedly, causing high computation cost and long computation time. Additionally, the iterative parameters need to be tuned delicately to seek for the desired image quality. 

Very recently, the deep learning (DL) based medical image reconstruction algorithms have emerged, and attempts have been made to perform DECT image reconstruction. So far, the DL-based DECT decomposition method published in literature are implemented in either the projection domain or the CT image domain. For instance, Zhang \textit{et al.} designed a Butterfly network to perform two material decomposition in the image domain\cite{Zhang2019ImageDD}. Chen \textit{et al.} proposed a VGG-loss based CNN with enlarged receptive field to obtain the decomposition results \cite{Chen2019RobustMD}. Clark \textit{et al.} used a U-net structure CNN to learn the material-specific maps from spectral CT images \cite{Clark2018MultienergyCD}. Wu \textit{et al.} proposed a fully convolutional DenseNet for the same purpose \cite{Wu2019MultimaterialDO}. Zhao \textit{et al.} developed a network to predict the high-energy CT image from the low-energy CT image \cite{Zhao2020ObtainingDC}. Some other studies also incorporate the CNN within the iterative decomposition framework to learn the convolutional regularizers \cite{Li2020ImageDomainMD} and the sparsifying transforms \cite{Li2020DECTMULTRADC} in image domain. On the other hand, Shi \textit{et al.} used a modified U-net to obtain the material-specific projections \cite{Shi2019RawDataBasedMD} in the projection-domain. Xu \textit{et al.} proposed a projection decomposition network to learn a compact spectrum representation \cite{Xu2018ProjectionDA}. Though better image quality have been achieved by the image-domain or projection-domain DL-based DECT material decomposition methods, a major drawback is that only single-domain information is considered, meaning that the prior information in the other domain has been underutilized. To the best of the authors' knowledge, there is no end-to-end DL network yet to take advantage of such mutual-domain knowledge in DECT imaging. 

In order to improve the DECT image quality and decomposition accuracy, we propose a \underline{d}oma\underline{i}n-t\underline{r}ansformation based network for \underline{e}nd-to-end material de\underline{c}omposi\underline{t}ion (DIRECT-Net), which makes full use of the mutual-domain (sinogram and CT image) prior information. The DIRECT-Net has immediate accesses to both the dual-energy sinogram and CT image data. Specifically, a projection-domain subnetwork and an image-domain subnetwork are cascaded together via a middle domain-transformation module. This particular network architecture is inspired by the AUTOMAP  \cite{Zhu2018ImageRB}. Instead of using the fully-connected CNN layers to perform domain transformation, we adopt a user-defined back-projection-based operator to transform the sinogram into the CT image domain\cite{Ge2020ADAPTIVENETDC}. By doing so, the computational resource could be well controlled and drastically saved. Compared to the traditional direct material decomposition algorithms, the DIRECT-Net is able to alleviate the heavy computation burden and greatly reduce the computation time. Most importantly, the DIRECT-Net has superior potentials to tolerate high image noise and spectral uncertainties by grasping the complicated prior information from a large amount of training data. In this work, our main contributions include: (1) Develop a new end-to-end network with mutual-domain knowledge for more accurate DECT material decomposition. (2) Propose a robust training data generation method that leverages the underlying physics of DECT imaging. The DIRECT-Net trained with those numerically simulated data could be directly utilized on experimental data, and generates accurate material-specific images.


The rest of this paper is organized as follows: Section \ref{sec: method} illustrates the DECT material decomposition model, the design of DIRECT-Net, the network implementation details, and the generation pipeline of the training data. Section \ref{sec: experiment} describes the experimental setup. Section \ref{sec: results} presents the results of numerical phantom, iodine solution phantom, and biological specimen. Section \ref{sec: conclusion} provides the discussions and a brief conclusion.

\section{Methods} \label{sec: method}
\subsection{Direct material decomposition model}
	\begin{figure*}[!t]
	\centering
	\includegraphics[width=1\textwidth]{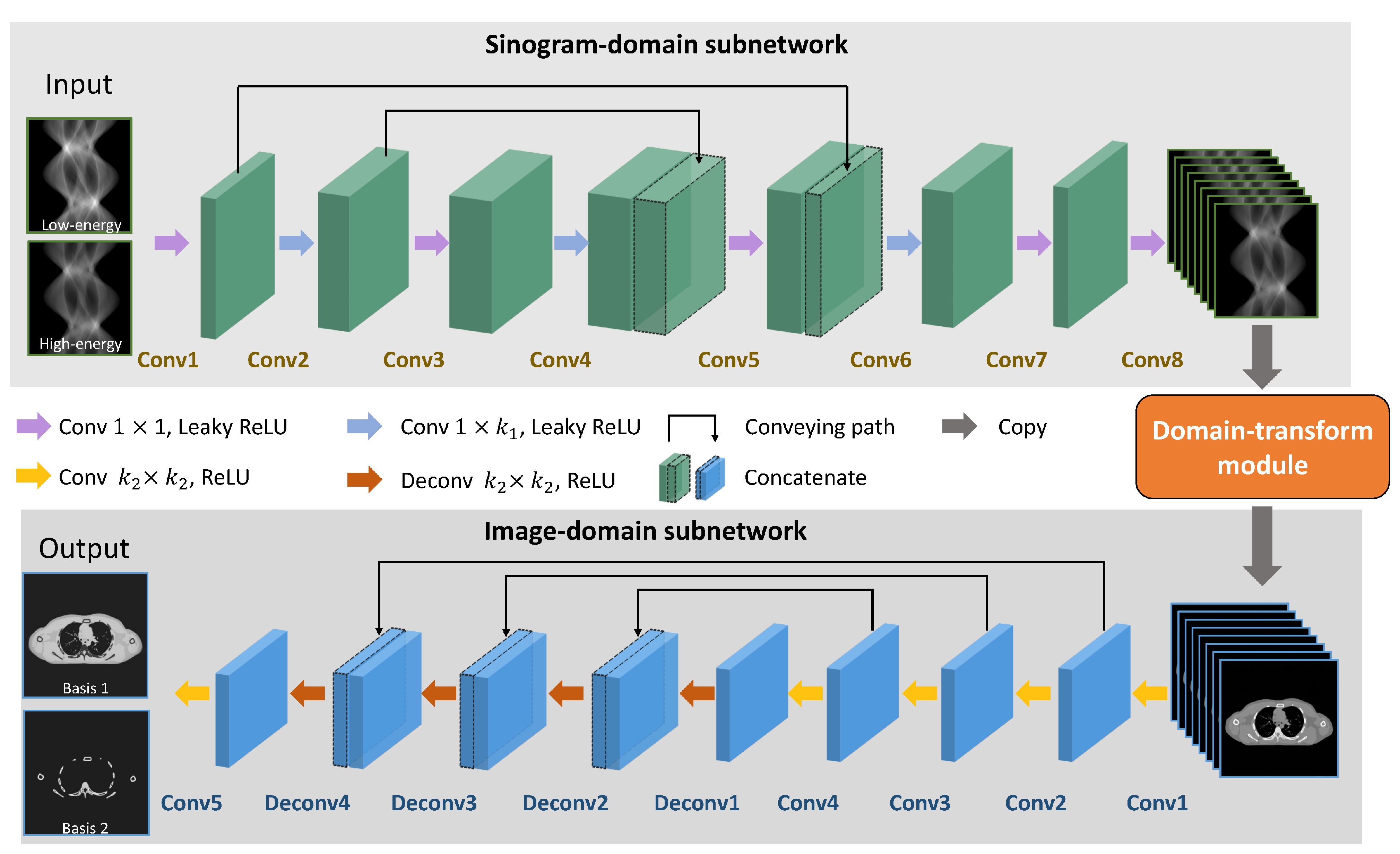}
	\caption{Architecture of DIRECT-Net. The main compartments include: the sinogram-domain subnetwork, the domain-transform module, and the image-domain subnetwork. . Dual-energy sinograms are taken as network input and the decomposed material-specific images are output. }
	\label{fig_network}
	\end{figure*}
The basic principle of material decomposition in spectral CT imaging lies in the fact that every material has a unique attenuation response to X-ray photons with different energies.
Such attenuation response is known as the linear attenuation coefficient $\mu(E, \overrightarrow{x})$, which is a product of the energy-dependent mass attenuation coefficient $\mu_{\rm{m}}(E)$ \cite{Hubbell1995TablesOX} and the location-dependent density distribution $\rho(\overrightarrow{x})$. The $\mu(E, \overrightarrow{x})$ of the object can be regarded as contributions of selected basis materials $\alpha$:
	\begin{equation}
	\mu(E, \overrightarrow{x})=\sum_\alpha \mu_{\rm{m}\alpha}(E) \rho_\alpha(\overrightarrow{x}).
	\label{eq1}
	\end{equation}

Let $I^{0}_{i,k}$ denotes the incident X-ray intensity for ray $i$ and spectrum $k$. Then, the beam intensity after crossing a scanned object can be expressed as:
	\begin{equation}
	I_{ik} = I^{0}_{ik} \int_E s_{ik}(E) {\rm exp} \left [ -\int_{t\in l_i} \mu(E, \overrightarrow{x}(t)) dt \right ] dE,
	\label{eq2}
	\end{equation}
where $s_{ik}(E)$ denotes the normalized X-ray beam intensity, $ \int_{t\in l_i}$ represents the integration along the $i^{\rm th}$ ray path $l_i$. In a discrete form, the object density map $\rho_\alpha$ is assumed to have $J$ pixels, and $\rho_{\alpha j}$ represents the density value of pixel $j$. By substituting \eqref{eq1} into \eqref{eq2}, we have the estimated measurement $\hat I_{ik}$:
	\begin{equation}
	I_{ik} \approx \hat I_{ik} = I^{0}_{ik} \sum_{e} s_{ike}{\rm exp}  \left [- \sum_{\alpha,j} \mu_{\rm{m}\alpha e} a_{ikj}  \rho_{\alpha j}\right ],
	\label{eq3}
	\end{equation}
where $a_{ikj}$ is the intersection length of ray $i$ with pixel $j$ for measurements with spectrum $k$. All the elements $a_{ikj}$ with different $i$ and $j$ compose the system matrix $A_k$. $A_k$ could be identical or distinct for measurements of different spectra, depending on the data acquisition strategy. 

For the purpose of estimating the material-specific density map $\rho$ from measurements $y_{ik}$, minimization of the constraint objective function is required. A general form is as follows:
	\begin{equation}
	\rho^* = \arg \min_\rho \left \{\sum_{i,k} D(\hat I_{ik}, y_{ik})  + \sum_\alpha R(\rho_\alpha)  \right \},
	\label{eq4}
	\end{equation}
where $D(\cdot)$ denotes the data consistency term, in which the statistical noise property could be incorporated; $R(\cdot)$ denotes the regularization term that considers various prior information such as TV, non-negativity, etc.

\subsection{DIRECT-Net}
Fig.~\ref{fig_network} shows the architecture of the proposed DIRECT-Net. It consists of a sinogram-domain subnetwork (SD-SubNet), a domain-transformation module (DT-module), and an image-domain subnetwork (ID-SubNet). Details of each compartment are discussed below:

\subsubsection{SD-SubNet}
As shown in Fig.~\ref{fig_network} (top), the SD-SubNet contains 8 convolution layers. The convolution kernels used in the subnetwork have two dimensions: $1\times 1$ and $1\times k_1$. The $1\times 1$ filters are employed to incorporate the pixel-wise spectral information between sinograms of different spectra for potential material decomposition in sinogram domain. The $1\times k_1$ filters are implemented along the detector row direction to partially denoise the sinogram. Moreover, it is also possible to ameliorate the Ramp filter that will be used by the DT-module for domain transformation. For this purpose, Leaky ReLU activation is used to allow small negative values when the input is less than zero. Each layer consists of $l$ filters with stride of 1. The conveying path, which copies the output of an earlier convolution layer to a later layer, is used. By concatenating these feature maps, image spatial resolution can be preserved while maintaining the network convergence \cite{Shan20183DCE}. In all, the SD-SubNet plays multiple roles of signal denoising, spectra augmentation, filter amelioration, and material decomposition. 

\subsubsection{DT-module}
To transform the sinogram data into the CT images, we build a DT-module based on analytical domain-transformation. In particular, the most widely used analytical filtered backprojection (FBP) reconstruction algorithm is integrated into the DIRECT-Net. Mathematically, the FBP reconstruction procedure is expressed as:
	\begin{equation}
	x = \mathbb{BP}(g*y),
	\label{eq5}
	\end{equation}
where $x\in\mathbb{R}^{n}$ represents the CT image, $y\in\mathbb{R}^{m}$ represents the sinogram, $g$ denotes the filter, and $\mathbb{BP}(\cdot)$ denotes the backprojection operation.

To make the domain transformation from $y$ to $x$ in DIRECT-Net compatible with the TensorFlow platform, the gradient back propagation also needs to be considered:
	\begin{equation}
	grad(x) = \mathbb{FP}(G^Ty),
	\label{eq6}
	\end{equation}
where the linear matrix $G$ corresponds to the convolution $g$ procedure and $\mathbb{FP}(\cdot)$ denotes the forward projection. We have developed the corresponding forward projection and backprojection operators based on the work of Gao \cite{Gao2012FastPA}. Both the GPU accelerated $\mathbb{BP}(\cdot)$ and $\mathbb{FP}(\cdot)$ operators take a series of system parameters as input and calculate the backprojection and forward projection in parallel by CUDA. Eventually, the DT-module can transform $K$ sinograms to $K$ CT images.

\subsubsection{ID-SubNet}

The structure of ID-SubNet is shown in Fig.~\ref{fig_network} (bottom). It takes the $K$ number of CT images from the DT-module as input, and finally outputs the decomposed material-specific density maps. Inspired by the CT image denoising works \cite{Shan20183DCE, Chen2017LowDoseCW, Zhang2018ASC} and the image-domain material decomposition network \cite{Clark2018MultienergyCD}, we construct the ID-SubNet with convolution layers, deconvolution layers and conveying paths. Each layer uses $l$ filters with kernel size of $k_2\times k_2$, followed by the ReLU activation. The stride is set to 1 and no pooling layer is used to avoid loss of image resolution. The stacked convolution layers (Conv1 to Conv4) are designed to extract features from low-level to high-level step by step. During this process, spacial prior information of the neighboring pixels within the network receptive field as well as the spectral prior information of multiple channels can be incorporated simultaneously. The stacked deconvolutional layers (Deconv1 to Deconv4) are used to recover image details from the extracted features. The last convolutional layer (Conv5) is applied to generate the output images with $M$ channels, where $M$ equals to the number of basis materials.

\subsection{Network parameter selection}
The parametric structure of all layers in the DIRECT-Net is shown in Table. \ref{table_1}. For the SD-SubNet, the $1\times1$ and $1\times3$ ($k_1=3$) filters are used alternatively from Conv1 to Conv6, and two $1\times1$ filters are followed in Conv7 and Conv8 in SD-SubNet. The number of filters $l$ varies from 8 to 128. The ID-SubNet contains 5 convolution layers and 4 deconvolution layers. Each layer uses 32 filters with kernel size of $3\times3$ ($k_2=3$), except for that only 2 filters are used in the final layer to guarantee the same channel number as of the two material bases. 

	{\renewcommand{\arraystretch}{0.9}
	\begin{table}[t]
	\caption{Parametric structure of each layer in the DIRECT-Net. The DT-module is a user-defined domain transformation operator, thus it contains no learnable parameters. }
	\label{table_1}
	\begin{tabular}{|c|c|c|c|}
	\hline
	Layer   & Parameters                 & No. of filters & No. of features \\ \hline
	\multicolumn{4}{|c|}{SD-SubNet}                                         \\ \hline
	Conv1   & conv 1$\times1$, Leaky ReLU & 32             & 32              \\ \hline
	Conv2   & conv 1$\times3$, Leaky ReLU & 64             & 64              \\ \hline
	Conv3   & conv 1$\times1$, Leaky ReLU & 128            & 128             \\ \hline
	Conv4   & conv 1$\times3$, Leaky ReLU & 128            & 128+64          \\ \hline
	Conv5   & conv 1$\times1$, Leaky ReLU & 96             & 96+32           \\ \hline
	Conv6   & conv 1$\times3$, Leaky ReLU & 64             & 64              \\ \hline
	Conv7   & conv 1$\times1$, Leaky ReLU & 32             & 32              \\ \hline
	Conv8   & conv 1$\times1$, Leaky ReLU & 8              & 8               \\ \hline
	\multicolumn{4}{|c|}{DT-module}                                         \\ \hline
	\multicolumn{4}{|c|}{ID-SubNet}                                         \\ \hline
	Conv1   & conv 3$\times3$, ReLU       & 32             & 32              \\ \hline
	Conv2   & conv 3$\times3$, ReLU       & 32             & 32              \\ \hline
	Conv3   & conv 3$\times3$, ReLU       & 32             & 32              \\ \hline
	Conv4   & conv 3$\times3$, ReLU       & 32             & 32              \\ \hline
	Deconv1 & deconv 3$\times3$, ReLU     & 32             & 32+32           \\ \hline
	Deconv2 & deconv 3$\times3$, ReLU     & 32             & 32+32           \\ \hline
	Deconv3 & deconv 3$\times3$, ReLU     & 32             & 32+32           \\ \hline
	Deconv4 & deconv 3$\times3$, ReLU     & 32             & 32              \\ \hline
	Conv5   & conv 3$\times3$, ReLU       & 2              & 2               \\ \hline
	\end{tabular}
	\end{table}}

\subsection{Training data preparation} \label{training_data}
\begin{figure}[htb]
	\centering
	\includegraphics[width=0.75\textwidth]{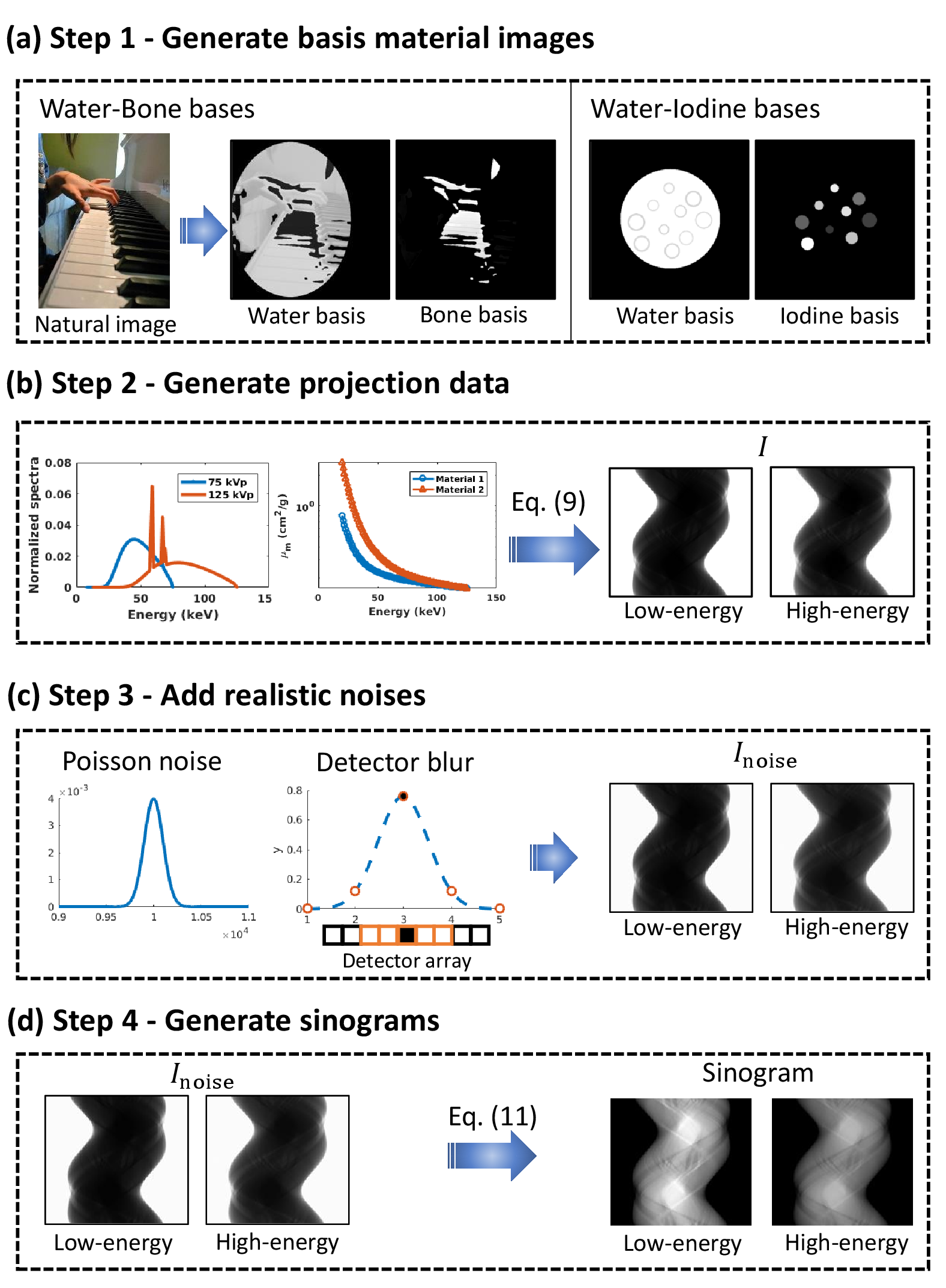}
	\caption{Key procedures to numerically generate the training data: (a) generate basis material images,  (b) generate projection data, (c) add realistic noises, and (d) obtain sinograms. }
	\label{fig_training_data}
	\end{figure}

In practice, the dual-energy projection data can be acquired readily from the commercial CT systems. However, it is usually challenging to get the corresponding ground truth material-specific density maps. Although the available decomposition algorithms provided by vendors are feasible to generate labels, but they still can not be considered as the ground truth basis images. Therefore, we propose to generate high quality training data based on physics-informed numerical simulation. As shown in Fig.~\ref{fig_training_data}, key steps include: 1) generate basis material images, 2) generate projection data, 3) add realistic noises, and 4) generate sinograms. The following content will introduce the detailed procedures. 

\subsubsection{Generate basis material images}
To create the training and testing DECT datasets, basis material images $\rho_\alpha$ need to be generated firstly. In this paper, 
we propose two strategies for the generation of $\rho_\alpha$. In case where the scanned sample contains fine patterns, natural images are used to generate the corresponding basis images, since they are rich in diverse details and a huge amount of data can be collected easily from the publicly accessible online database ImageNet \cite{5206848}. On the other hand, when the structure of scanned sample is relatively simple, we construct the basis images directly through numerical simulation. Considering the two physical experiments that will be introduced in Section \ref{sec: physical_experiment}, we use the first strategy to create the Water-Bone bases, and the second strategy for Water-Iodine bases. 

\begin{enumerate}
\item[-] Water-Bone bases. A huge number of natural images downloaded from the ImageNet are used to generate the Water-Bone basis labels. First, a fixed channel of each RGB image, the R channel for example, is extracted and resized to $512\times512$. This image is used as the original image $I_{\rm org}$. Second, a smoothing function $\mathcal F $ is applied to the original image to create an ``index image". The basis masks $M_\alpha$ (bone mask and water mask) are generated by threshold segmentation $\mathcal{T}_\alpha$ of the ``index image":
	\begin{equation}
	M_\alpha = \mathcal T_\alpha(\mathcal F (I_{\rm org})),
	\label{eq7}
	\end{equation}

Afterwards, the original image is multiplied with each of the individual masks, and the image values are adjusted to be consistent with practical densities. Finally, the images are multiplied by an ellipse mask $M_{\rm elp}$ with randomly varied long and short axes, and so far basis images $\rho_\alpha$ are obtained:

	\begin{equation}
	\rho_\alpha = \mathcal S_\alpha(I_{\rm org} \cdot M_\alpha) \cdot M_{\rm elp},
	\label{eq8}
	\end{equation}

where $\mathcal S_\alpha$ represents the value adjust operator. An example of the generated basis images is shown in the left part of Fig.~\ref{fig_training_data} (a). In particular, the density of water basis varies from 0.85~g/cm$^3$ to~1.15 g/cm$^3$, and the density of bone basis varies from 1.19~g/cm$^3$ to 1.61~g/cm$^3$. Both basis images have overlapping regions, 
whose values are set relatively lower from 0 to 0.15~g/cm$^3$.
\item[-] Water-Iodine bases. These basis material images are generated directly through numerical simulations. Digital phantoms containing 10 iodine inserts with diameters varying from 9~mm to 18~mm are simulated. The inserts are randomly positioned within the simulated water tank, whose diameter also changes randomly. Moreover, the concentrations of iodine solutions vary from 0 to 30~mg/cc. An example of the generated water and iodine basis images is shown in the right part of Fig.~\ref{fig_training_data} (a).
\end{enumerate}

\subsubsection{Generate projection data} 
With the obtained basis material density maps $\rho_\alpha$, the projection data $I$ can be calculated readily using \eqref{eq3}, which is recalled here:
	\begin{equation}
	I_{ik} \approx\hat I_{ik} = I^{0}_{ik} \sum_{e} s_{ike}{\rm exp}  \left [- \sum_{\alpha,j} \mu_{\rm{m}\alpha e} a_{ikj}  \rho_{\alpha j}\right ].
	\label{eq9}
	\end{equation}

Specifically, the normalized X-ray spectra $s_{ike}$ of the low- and high-energy are generated using the SpekCalc \cite{Poludniowski_2009} according to the experimental setup of data acquisition, the mass attenuation coefficients $\mu_{\rm{m}\alpha e}$ of basis materials are obtained from the NIST database \cite{Hubbell1995TablesOX}, see the plotted curves in Fig.~\ref{fig_training_data} (b). The total incident X-ray intensity is set to $2\times10^6$. The system parameter $a_{ikj}$ is considered by our user-defined forward projection operator $\mathbb{FP}(\cdot)$. In particular, the geometry parameters required by $\mathbb{FP}(\cdot)$ are set to be the same as those used for experiment data acquisition. Additionally, the CT image pixel size is set to 0.35~mm $\times$ 0.35~mm. An example of the generated projection data is shown in the right part of Fig.~\ref{fig_training_data} (b).

\subsubsection{Add realistic noises}
In reality, CT imaging encounters different kinds of noises, including the quantum Poisson noise, electronic Gaussian noise, etc. To mimic the real detector response, we simulate the quantum noise and the electronic noise with a Gaussian distributed noise model $\mathcal N(p, \sigma)$, where the mean value $p$  is set as the expected X-ray intensity, and the variance is set as $\sigma = \beta\times p$. The calibration factor $\beta$ is determined from experiments. Additionally, noise correlation effect is simulated by filtering the projection data with a 1-dimension (1D) Gaussian kernel $g_d$ along the detector direction. The kernel size and standard deviation are defined by comparing with the experimentally measured noise power spectrum (NPS) curves. Eventually, the noisy projection data $I_{\rm noise}$ is expressed as:
	\begin{equation}
	I_{\rm noise} = g_d*(I + \mathcal N(p, \sigma)). 
	\label{eq10}
	\end{equation}
In our experiments, the measured $\beta$ for low-energy and high-energy are 1.2 and 1.7, respectively; the kernel size of $g_d$ is $1\times5$ and the standard deviation is 0.52.
\subsubsection{Generate sinograms} 
According to \eqref{eq9}, the sinogram is obtained as:
	\begin{equation}
	{\rm sino} = -\ln(I_{\rm noise}/I^0).
	\label{eq11}
	\end{equation}
These obtained sinograms and the basis material CT images are used as network input and labels, respectively. For each experiment, the dataset finally consists of 10000 training samples and 100 validation samples.

\subsection{Network Training Details}\label{training_detail}
The proposed DIRECT-Net contains a set of parameters $\Theta$ that need to be optimized. In this work, the mean squared error (MSE) between the network estimated material density $\hat\rho_\alpha(\Theta)$ and the label density $\rho$ is used as the loss function. Explicitly, it is defined below:
\begin{equation}
	\mathcal L(\Theta) =\frac{1}{MN} \sum_\alpha \lambda_\alpha \left\| \hat\rho_\alpha(\Theta) - \rho_\alpha \right \|_F^2,
	\label{eq11}
	\end{equation}
where $M$ and $N$ are the numbers of CT image pixels along the vertical and horizontal dimension, $\left \| \cdot \right \|_F^2$ denotes the Frobenius norm, $\lambda_\alpha$ is a factor to balance the loss weight of different materials. We chose the parameter $\lambda_\alpha$ according to the ratio of mean squares of basis materials in the training dataset.

The DIRECT-Net was trained on TensorFlow platform with a single NVIDIA GeForce GTX 1080Ti GPU card. The Adam optimization algorithm was used with a starting learning rate of $10^{-4}$, which exponentially decayed by a factor of 0.98 after every 500 steps. The network was trained for 20 epochs for each experiment data. Individual training process took about 121 hours.


\subsection{Comparison Algorithms}
To evaluate the performance of our newly proposed DIRECT-Net, three different material decomposition methods are compared. 

\subsubsection{ID-EP algorithm} This algorithm is a state-of-the-art image-domain algorithm proposed by Niu \textit{et al.} \cite{Niu2014IterativeID}. It uses an edge-preserving regularizer. Specifically, an effective mass attenuation coefficient $\mu_{\rm m}$ needs to be decided for each basis material, which has a large influence on the decomposition results. In this study, $\mu_{\rm m}$ is selected by measuring the mean values within certain regions of interest (ROIs) in the dual-energy CT images. For example, the ROI of $\mu_{\rm m}$ for water is selected on the center of the iodine solution CT images (i.e. water tank region), and the ROI of $\mu_{\rm m}$ for bone is selected on the pure bone region. 

\subsubsection{Direct-JSI algorithm} This algorithm is a direct decomposition algorithm proposed by Mechlem \textit{et al.}  \cite{Mechlem2018JointSI}. It starts with fine-tuning the forward-projection model with multiple calibration measurements and then solves the negative log-likelihood objective function iteratively based on separable surrogate functions. In this study, such fine-tune step is omitted for lack of calibration experiments. The X-ray spectra and material attenuation coefficients required for algorithm implementation are from the SpekCalc software and NIST database.

\subsubsection{ID-Net} This method corresponds to a pure image-domain network, which has the same structure as the ID-SubNet of the proposed DIRECT-Net. The ID-Net takes the FBP reconstructed dual-energy CT images as the input. The training data are generated following the same procedure discussed in \ref{training_data}.

For both ID-EP and Direct-JSI methods, the iteration parameters are carefully selected manually to achieve the best performance.

\section{Experiments} \label{sec: experiment}
\subsection{XCAT simulation experiment}
The extended cardiac-torso (XCAT) phantom \cite{Segars2008RealisticCS} was used to verify the material decomposition performance of DIRECT-Net trained with the Water-Bone bases dataset prepared using natural images. A slice of $512\times512$ image was extracted from the XCAT phantom. With threshold segmentation and value adjustment, the bone and water basis material images were obtained, as shown in Fig.~\ref{fig_xcat_image}(a). Corresponding sinogram data were then generated following the procedure in \ref{training_data}.

\subsection{Physical experiments} \label{sec: physical_experiment}
The experiment data were acquired on an in-house cone-beam CT imaging system in our lab. The system is equipped with a rotating-anode Tungsten target diagnostic grade X-ray tube (Varex G-242, Varex Imaging Corporation, UT, USA). The low-energy data were obtained under 75~kVp with 1.5~mm aluminum (Al) and 0.2~mm copper (Cu) filtration. The high-energy data were obtained at 125 kVp, with 1.5~mm Al and 1.2~mm Cu filtration. The flat panel detector (Varex 4343CB, Varex Imaging Corporation, UT, USA) has $3072\times3072$ pixels with a native dimension of 0.139~mm $\times$ 0.139~mm. During the data acquisition, the detector was operated at the $3\times3$ binning mode. The source to detector distance was 1560.6~mm, and the source to rotation center distance was 1156.3~mm. Projection data were collected from the angular range of $360^{\circ}$ with intervals of $0.4^{\circ}$.

	\begin{figure}[!h]
	\centering
	\includegraphics[width=0.75\textwidth]{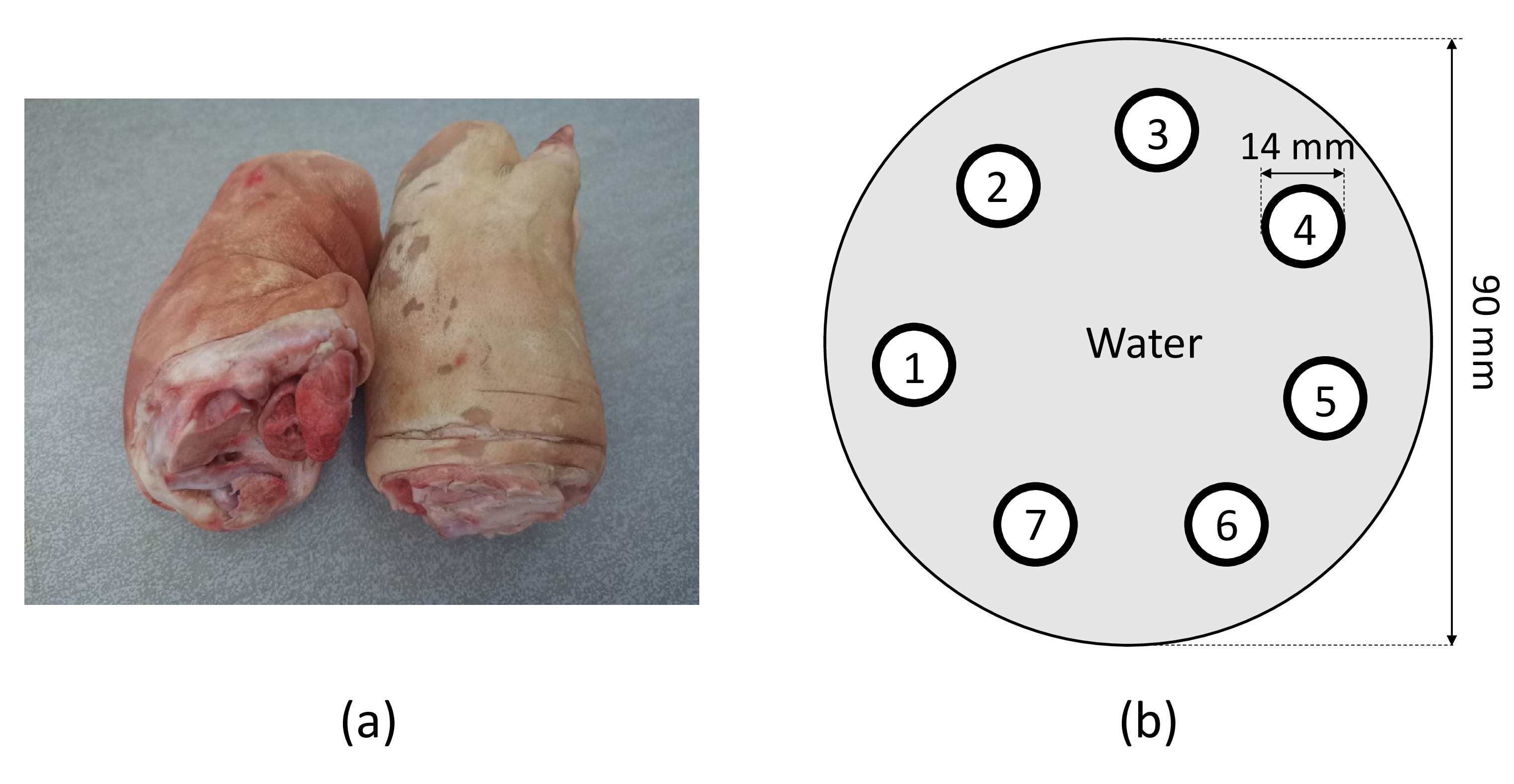}
	\caption{(a) The pig knuckle specimen. (b) Schematic diagram of the iodine solution phantom, in which the iodine concentration in inserts 1 to 7 are 2.0~mg/cc, 2.5~mg/cc, 5.0~mg/cc, 7.5~mg/cc, 10.0~mg/cc, 15.0~mg/cc and 20.0~mg/cc, respectively.}
	\label{fig_samples}
	\end{figure}

Two samples were scanned in this study. The first is a pig knuckle specimen, and the second sample is a iodine solutions phantom with different concentrations, see Fig.~\ref{fig_samples}. Specific acquisition parameters are listed below:

\subsubsection{Pig knuckle specimen}
For this experiment, the X-ray tube current was set at 8 mA and 5 mA for the low-energy and high-energy spectra, respectively. 

\subsubsection{Iodine solution phantom}
We prepared iodine solutions of different concentrations (2~mg/cc to 20~mg/cc) by diluting the iobitridol contrast agent (Xenetix 350, Guerbet company, France).
The solutions were then filled in 7 plastic tubes with diameter of 14~mm. All tubes were emerged in a cylinder water tank with diameter of about 90~mm. 
The X-ray tube current was set at 10.2~mA and 7.1~mA for low-energy and high-energy spectra, respectively.  

\section{Results} \label{sec: results}
\subsection{XCAT simulation results}

\begin{figure*}[!t]
	\centering
	\includegraphics[width=1\textwidth]{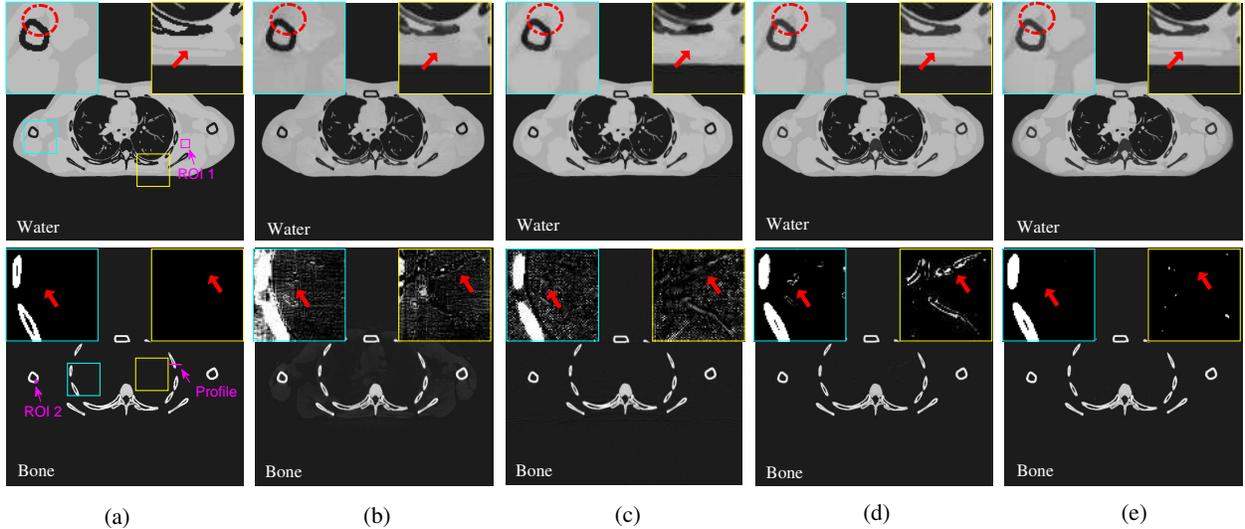}
	\caption{The decomposition results of the XCAT phantom: (a) label, (b) ID-EP, (c) Direct-JSI, (d) ID-Net, (e) DIRECT-Net. The
top and bottom rows show the water and bone basis images with display window of [-0.2, 1.5]~g/cm$^3$. For exception, the display window of the magnified ROIs in the second row is [0, 0.07]~g/cm$^3$. }
	\label{fig_xcat_image}
	\end{figure*}

The validation results of the XCAT phantom are shown in Fig.~\ref{fig_xcat_image}(b) to (e). Images in each column represent the decomposition results using a certain method. It can be observed that the decomposition performance varies depending on individual methods. Specifically, the water basis images generated from the ID-EP and Direct-JSI methods have less satisfied spatial resolution: the fine muscle border denoted by red arrows and the minor features denoted by red circles become invisible, see Fig.~\ref{fig_xcat_image} (b) and (c). On the contrary, those features are well maintained on the images obtained from the ID-Net and DIRECT-Net. For the decomposed water basis, the DIRECT-Net shows slightly better performance than the ID-Net. Regarding the decomposed bone basis images, the ID-EP, Direct-JSI and ID-Net methods generate a lot of residual artifact signals outside of the bone regions, as marked by the red arrows in the magnified ROI images. Clearly, those artifacts are well suppressed in Fig.~\ref{fig_xcat_image}(e) (bottom) by DIRECT-Net, indicating its superior material decomposition performance.
	
	\begin{figure}[!t]
	\centering
	\includegraphics[width=0.85\textwidth]{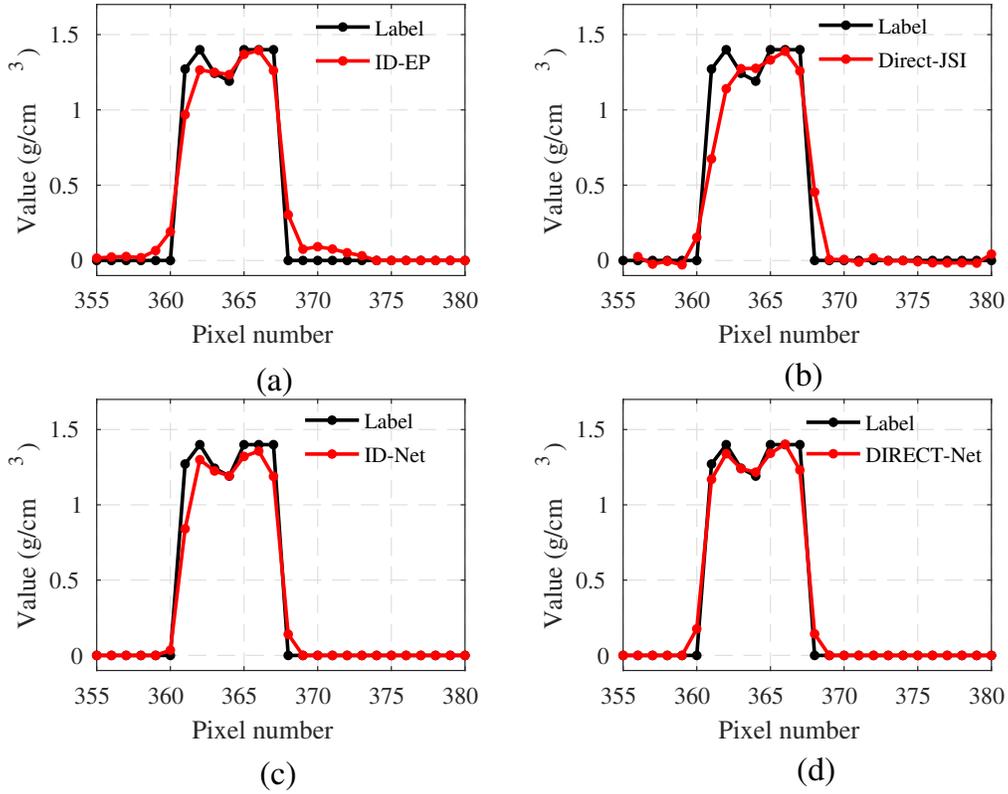}
	\caption{Line profile comparison results along the highlighted magenta region in Fig.~\ref{fig_xcat_image}: (a) ID-EP, (b) Direct-JSI, (c) ID-Net, (d) DIRECT-Net. }
	\label{fig_xcat_profile}
	\end{figure}

The mean value and the standard deviation (STD) of ROI~1 (muscle region) and ROI~2 (bone region) highlighted on Fig.~\ref{fig_xcat_image}(a) are analyzed quantitatively. Results are listed in Table.~\ref{table_2}. As shown, the ID-EP method produces the largest STD, indicating the highest noise level. Whereas, the proposed DIRECT-Net provides the highest accuracy. In addition, line profile (pixels denoted by the magenta line in Fig.~\ref{fig_xcat_image}(a)) comparison results are shown in Fig.~\ref{fig_xcat_profile}. Clearly, the DIRECT-Net well preserves the object details with the highest precision.
	{\renewcommand{\arraystretch}{1.4}
	\begin{table}[!h]
	\caption{The decomposed material densities {\upshape (g/cm$^3$)} for different methods. ROI~1 and ROI~2 correspond to the muscle region and the bone region, respectively, see Fig.~\ref{fig_xcat_image}. }
	\label{table_2}
	\centering
\begin{tabular}{ccc}
\hline  \hline
\multicolumn{1}{l}{} & ROI 1        & \multicolumn{1}{c}{ROI 2}        \\ \hline
Truth                & 1.050        & 1.383                            \\
ID-EP            & 0.978$\pm$0.0202 & 1.295$\pm$0.1170                     \\
Direct-JSI        &1.054$\pm$0.0053 & \multicolumn{1}{l}{1.285$\pm$0.0649} \\
ID-Net            & 1.035$\pm$0.0034 & 1.286$\pm$0.0872                     \\
DIRECT-Net           & \textbf{1.046$\pm$0.0047} & \textbf{1.311$\pm$0.0753}  
\\ \hline  \hline      
\end{tabular}
	\end{table}}

\subsection{Pig knuckle experiment results}
	\begin{figure*}[!t]
	\centering
	\includegraphics[width=1\textwidth]{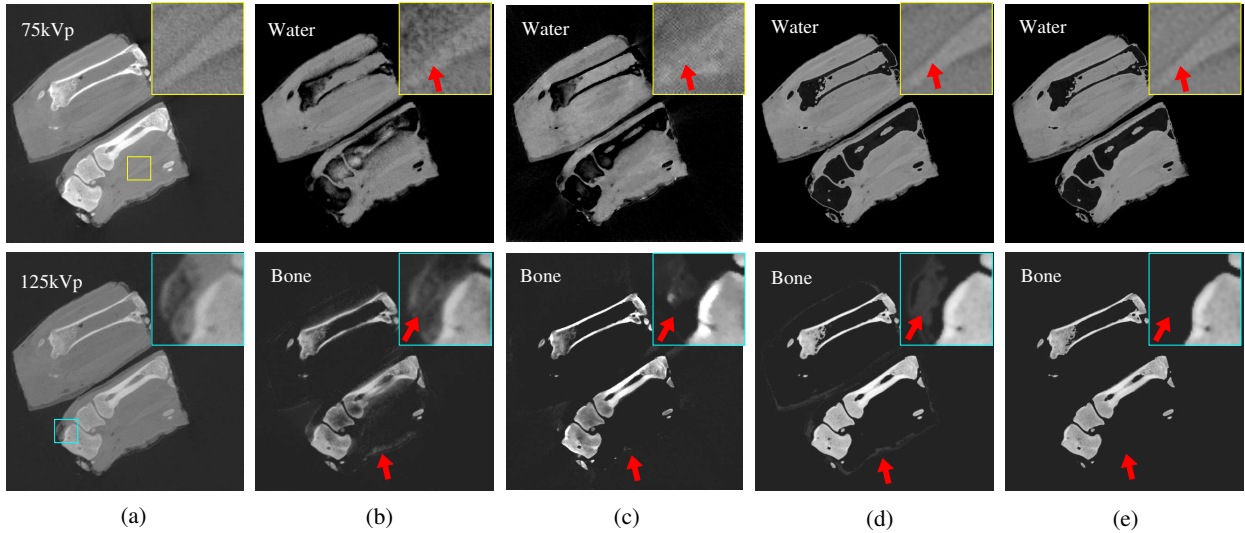}
	\caption{The DECT imaging results of the pig knuckle specimen. The low-energy and high-energy CT images are presented in (a) with a display window of [-0.1, 0.5]~cm$^{-1}$. Images in (b) to (e) are the decomposed basis results obtained from: ID-EP, Direct-JSI, ID-Net, and DIRECT-Net, correspondingly. The display window for the water basis images in top is [0.5, 1.5]~g/cm$^3$, and is [-0.3, 1.8]~g/cm$^3$ for the bone basis in bottom.}
	\label{fig_pig_image}
	\end{figure*}
	
\begin{figure*}[!h]
\centering
\includegraphics[width=1\textwidth]{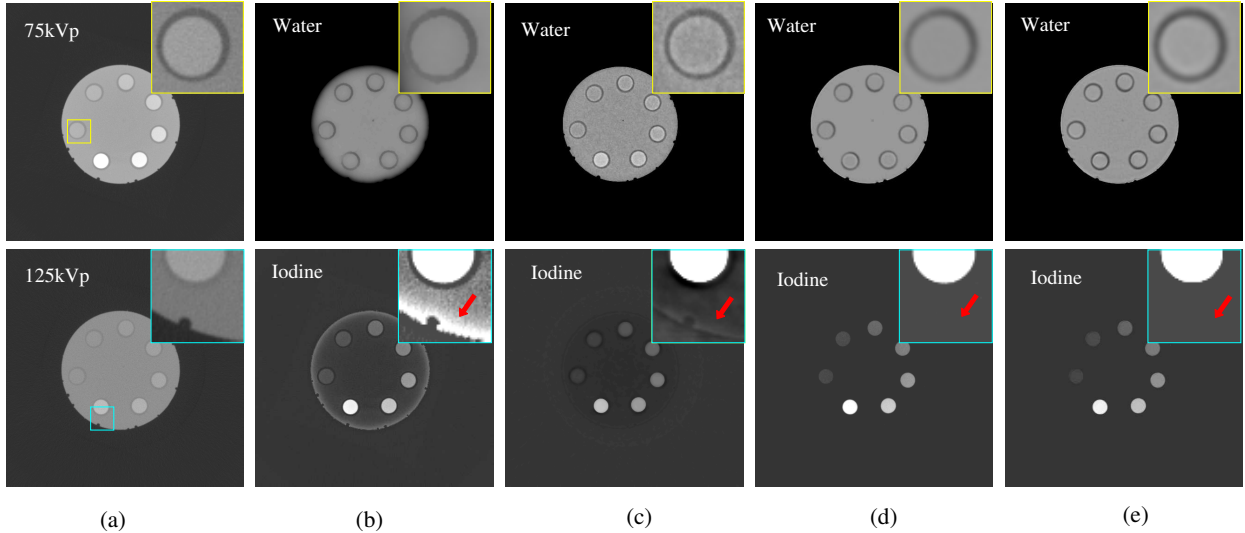}
\caption{The DECT imaging results of the iodine solution phantom. The low-energy and high-energy CT images are presented in (a) with a display window of [-0.1, 0.4]~cm$^{-1}$. Images in (b) to (e) are the decomposed basis results obtained from: ID-EP, Direct-JSI, ID-Net, and DIRECT-Net, correspondingly. The display window for the water basis images in top is [700, 1200]~mg/cc, and is [-6, 22]~mg/cc for the bone basis in bottom.}
\label{fig_iodine_image}
\end{figure*}

	{\renewcommand{\arraystretch}{1.1}
	\begin{table}[htb]
	\caption{The estimated mean values ($\mu$), standard deviations ($\sigma$), and relative errors of the iodine concentrations by different methods. }
	\label{table_3}
	\centering
\begin{tabular}{cccccccc}
\hline  \hline
Methods   & 2                  & 2.5                & 5                  & 7.5                & 10                 & 15                  & 20                   \\ \hline
\multirow{2}{*}{ID-EP}     & 3.10$\pm$0.36          & 3.44$\pm$0.38          & 6.06$\pm$0.38          & 8.31$\pm$0.39          & 10.77$\pm$0.41         & 15.31$\pm$0.36          & \textbf{20.43$\pm$0.45} \\
                                 & 55.12\%              & 37.72\%              & 21.24\%              & 10.77\%              & 7.70\%               & 2.08\%                &  \textbf{2.13}\%                \\
\multirow{2}{*}{Direct-JSI}   & 0.98$\pm$0.11          & 1.25$\pm$0.15          & 3.33$\pm$0.17          & 5.21$\pm$0.11          & 7.20$\pm$0.18          & 10.96$\pm$0.10          & 15.15$\pm$0.15          \\
                                 & -51.06\%             & -50.11\%             & -33.47\%             & -30.55\%             & -27.99\%             & -26.95\%              & -24.26\%              \\
\multirow{2}{*}{ID-Net}       & 2.30$\pm$0.31          & 2.78$\pm$0.33          & 5.59$\pm$0.32          & 8.14$\pm$0.35          & 10.79$\pm$0.35         & 16.25$\pm$0.34          & 22.17$\pm$0.39          \\
                                 & 15.10\%              & 11.01\%              & 11.86\%              & 8.59\%               & 7.94\%               & 8.30\%                & 10.85\%               \\
\multirow{2}{*}{DIRECT-Net}      & \textbf{1.94$\pm$0.20} & \textbf{2.51$\pm$0.19} & \textbf{5.17$\pm$0.20} & \textbf{7.53$\pm$0.20} & \textbf{9.95$\pm$0.21} & \textbf{14.86$\pm$0.21} & 20.46$\pm$0.26          \\
                                 &  \textbf{-3.05}\%              &  \textbf{0.57}\%               & \textbf{3.36}\%               &  \textbf{0.46}\%               &  \textbf{-0.51}\%              &  \textbf{-0.94}\%               & 2.28\%               \\ \hline  \hline
\end{tabular}
	\end{table}}

The decomposition results of the pig knuckle specimen are shown in Fig.~\ref{fig_pig_image}. In particular, images in Fig.~\ref{fig_pig_image}(a) are the FBP reconstructed low-energy and high-energy CT images. Fig.~\ref{fig_pig_image}(b) to (e) illustrate the decomposition results obtained from different methods. For the ID-EP method, the water basis image contains most of the tissues. However, the edge of the muscle structure, see the red arrow in the magnified ROI image, is contaminated by noises. Moreover, some bony structures are remained in the water basis image. Meanwhile, some tissue components also reside in the bone basis image, see the red arrows on the bottom image. Such cross-talk phenomenon could be caused by two reasons: the first one is the relatively high noise level in the reconstructed CT images; the second one is that the ID-EP algorithm is sensitive to the selected value of the attenuation coefficent $\mu_{\rm m}$ for basis materials. Our choice of $\mu_{\rm m}$ for bone may not be general enough since the intensity of bone varies heavily for different regions. In Fig.~\ref{fig_pig_image}(c), the separation between bone and tissues is better than ID-EP, but some tissue component still remains in the bone image, as highlighted by the red arrows. Another problem is the degraded image quality of the water basis image. The ``checkerboard artifacts" might be caused by the iteration procedure for the ill-posed maximum-likelihood reconstruction \cite{4307826}. Results in Fig.~\ref{fig_pig_image}(d) and (e) show that superior DECT image quality can be obtained from the network based methods. Take the zoomed-in tissue and bone regions as an example, their fine textures can all be well-preserved. Compared with the ID-Net (still contains slight cross-talk effects), the newly developed DIRECT-Net provides the most clean bone basis image with the fewest cross-talk effects. As a consequence, the mutual-domain information is important for high quality and accurate DECT imaging, and thus should be considered.

\subsection{Iodine solution experiment results}

To quantitatively evaluate the decomposition accuracy of the DIRECT-Net, we acquired experimental DECT data with iodine solutions of different concentrations. Fig.~\ref{fig_iodine_image}(a) shows the FBP reconstructed dual-energy CT images at 75~kVp and 125~kVp, respectively. Results in the Fig.~\ref{fig_iodine_image}(b) to (e) show the decomposed basis images using different methods. Visually, the ID-EP method has limited performance. Cupping artifacts in the decomposed iodine image are obvious, as highlighted by the red arrow. This is mainly caused by beam hardening effects, which exit in the FBP CT images and are further magnified by the decomposition procedure.
In Fig.~\ref{fig_iodine_image}(c), the Direct-JSI method reduces the beam hardening artifacts efficiently from projection domain, but some background noises are still significant. As a contrary, the ID-Net and DIRECT-Net generate the basis images with much better quality: water and iodine are well separated without cross-talk, noises and beam-hardening artifacts are also suppressed efficiently. From visual perception, the two methods have similar performance. 

	\begin{figure}[htb]
	\centering
	\includegraphics[width=0.85\textwidth]{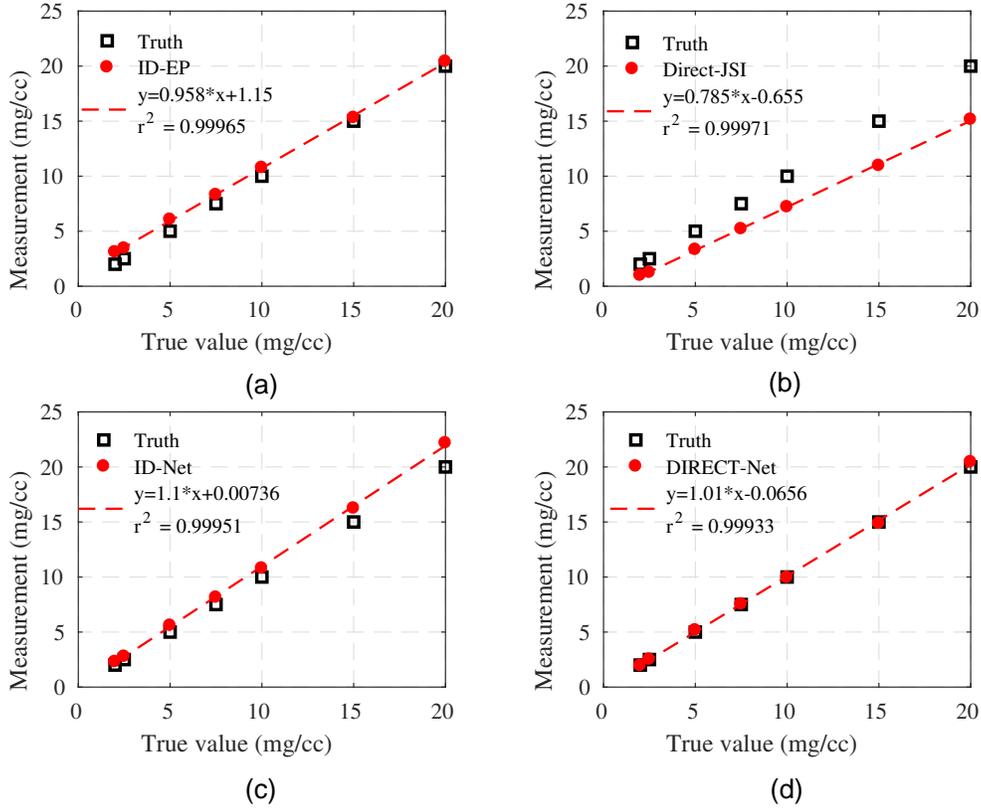}
	\caption{Plots of the measured iodine densities versus the true values for the different decomposition methods: (a) ID-EP, (b) Direct-JSI, (c) ID-Net, (d) DIRECT-Net. Linear fittings are illustrated for the experimental data.}
	\label{fig_iodine_profile}
	\end{figure}

We measured the mean value and STD of the pixels inside each iodine tube, as well as the relative errors of the mean values. The results are listed in Table. \ref{table_3}. For the wide range of iodine concentrations (from 2~mg/cc to 20~mg/cc), the proposed DIRECT-Net has quantification errors less than 4\%. It outperforms the ID-Net and other methods in almost all cases except for the 20 mg/cc iodine insert. The ID-EP method obtains the best result for 20~mg/cc iodine concentration. This is because that the $\mu_{\rm m}$ of iodine required by the ID-EP algorithm was measured on this area. Hence, this result in fact demonstrates a precise pre-calibration. In case no pre-calibrations were performed, the results obtained from the ID-EP method should become higher.  There is relatively big discrepancies between the quantification results obtained by the Direct-JSI method and the ground truth. One plausible cause is that we did not perform fine-tunes of the forward model as proposed by the authors. The ID-Net also shows inferior quantitative performance without assistance of the sinogram-domian information. 
Fig.~\ref{fig_iodine_profile} plots the measured iodine density versus the ground truth using different methods. It can be observed that the DIRECT-Net generates the most consistent results with the ground truth. Essentially, the linear fitting line in Fig.~\ref{fig_iodine_profile}(d) gets a slope of nearly 1 and a very small intercept, indicating the very good capability of DIRECT-Net in generating high accurate basis images even at very low iodine concentrations.

Finally, as for the computation time, in this specific case of water-iodine decomposition, the ID-EP method spent about 625 seconds for 25000 iterations to obtain images in Fig.~\ref{fig_iodine_image}; the Direct-JSI method took about 627 seconds for 40 iterations; the ID-Net used about 0.43 seconds and the DIRECT-Net need about 1.91 seconds.  After all, the network methods can greatly reduce the computation time for material decomposition, compared with conventional iterative algorithms.

\section{Discussions and Conclusion} \label{sec: conclusion}
With the purpose to improve the basis image quality and quantitative accuracy in DECT imaging, this study proposed a novel end-to-end mutual-domain material decomposition network (DIRECT-Net) based on deep learning technique. The performance of DIRECT-Net was evaluated by numerical and physical experiments. Both qualitative and quantitative results demonstrate that the DIRECT-Net is able to reduce image noise, improve signal accuracy, and save reconstruction-time.

This work was motivated by two simple observations. The first observation is that the one-step direct decomposition algorithms have great advantage in reducing artifacts and noises compared to the two-step image-domain or projection-domain algorithms \cite{Mechlem2018JointSI, Foygel_Barber_2016, Long2014MultiMaterialDU, Mory2018ComparisonOF}. However, this type of iterative algorithms usually require large computer memory, long computation time, difficult selections of parameters, and thus definitely should be improved. Second, we also noticed that deep CNN has powerful capabilities in medical image reconstructions, for example, reducing image noise, removing artifacts, and shortening the image reconstruction time \cite{8103129, 8340157}. Based on the above two observations, we developed a mutual-domain direct decomposition network that inherits the one-step DECT imaging idea but accelerates and improve the image reconstruction procedure elegantly.

In particular, this unified DIRECT-Net consists of a sinogram-domain subnetwork, a domain-transform module and an image-domain subnetwork. By design, the SD-SubNet plays a role of potential spectra augmentation and material decomposition, the DT-module transforms the sinogram data into image domain, and the ID-SubNet further performs image denoising and material decomposition. In addition, we have proposed a robust scheme to generate the training data from natural images through numerical simulation, rather than collecting them from real experiments. Results show that the proposed training data generation method is reliable, the DIRECT-Net trained with simulated data can be directly applied to the experimental data and accurate basis images can be obtained.

In spite of the promising outcome, this study has some potential room to be improved. First, the network structure can be improved. The SD-SubNet and ID-SubNet were designed empirically, especially the conveying paths and number of layers. Adjustments of the network may enhance the results, such as the convolution kernel sizes, the number of filters used in each layer and the network depth. According to our experiments, by increasing the number of output feature maps $K$ from the SD-SubNet, the performance of DIRECT-Net can be improved. In this work, we chose $K=8$ to make a balance between the performance and the network training time. Second, the network loss function can be modified. In our study, the loss weight of different materials is pre-selected according to the intensity (ratio of mean squares) of the basis images in the training dataset. This may not be rigorous enough since the network learns to constrain each image differently, and the learned MSE ratio of individual materials is not the same as our assumption. Therefore, fine-tunes could be helpful in obtaining better image quality. 

Based on the results presented in this article, a number of possible researches are of great interest for future study. For example, we focused on the two-material decomposition problem in this paper, however, the network can easily be adapted for multiple materials decomposition, which is required in other spectral CT imaging scenarios. This is of great importance for many clinical applications, such as the contrast enhanced liver-fat quantification \cite{Hyodo2017MultimaterialDA}, colonography, \cite{Muenzel2017SpectralPC}, atherosclerotic plaque imaging \cite{SiMohamed2017ReviewOA}, and so on. Therefore, it will be an interesting work to combine the present study with the rapidly developed multi-energy photon counting CT for characterization of multiple materials. Moreover, the concept of DIRECT-Net can be adopted for generating synthetic images from spectral CT, such as the virtual monochromatic images, the virtual non-contrast images, electron density and effective atomic number images \cite{McCollough2020PrinciplesAA}. The applications on other imaging modalities, such as the dual-energy mammography, dual energy digital breast tomosynthesis (DBT), etc. can also be investigated in future.

In summary, we have developed an end-to-end material decomposition network for quantitative DECT imaging. The proposed DIRECT-Net incorporates mutual-domain prior information to obtain high quality material-specific images under the deep learning framework. A robust and reliable training data generation scheme has been developed. Experimental validation results demonstrate that the DIRECT-Net provides an effective approach to suppress the noise magnification effect in DECT imaging, and accurate basis images can be generated with greatly reduced computation time. 

\section*{Acknowledgment}
The authors would like to acknowledge the ImageNet organization for providing images. The authors also would like to thank Dr. Zhanli Hu for providing the numerical XCAT phantom.

\bibliography{mybib}

\end{document}